\documentclass[aps,twocolumn,amsmath,amssymb,floatfix,reprint,footinbib,superscriptaddress,longbibliography,showkeys]{revtex4}
\usepackage{amsmath}
\usepackage{amscd}
\usepackage{enumitem}
\usepackage{mdwlist}
\usepackage{array}
\usepackage{graphicx}
\usepackage{pdfpages}
\usepackage{subfigure}
\usepackage{longtable}
\usepackage{appendix}
\usepackage{amsfonts}
\usepackage{multirow}
\usepackage{textcomp}
\usepackage{bm}
\usepackage{float}
\usepackage{url}
\usepackage[colorlinks]{hyperref}
\hypersetup{%
    plainpages=true,
    breaklinks=true,
    hypertexnames=false,
    pageanchor=true,
    bookmarksnumbered=true,
    colorlinks=true,
    linkcolor={blue},
    citecolor={red},
    urlcolor={blue},
    anchorcolor={black}
}

\begin{document}

\title{Fast and high-fidelity generation of steady-state entanglement using pulse modulation and parametric amplification}

\author{Ye-Hong Chen}
\affiliation{Theoretical Quantum Physics Laboratory, RIKEN Cluster for Pioneering Research, Wako-shi, Saitama 351-0198, Japan}
\author{Wei Qin}
\affiliation{Theoretical Quantum Physics Laboratory, RIKEN Cluster for Pioneering Research, Wako-shi, Saitama 351-0198, Japan}
\author{Franco Nori}
\affiliation{Theoretical Quantum Physics Laboratory, RIKEN Cluster for Pioneering Research, Wako-shi, Saitama 351-0198, Japan}
\affiliation{Department of Physics, University of Michigan, Ann Arbor, Michigan 48109-1040, USA}

\date{\today}

\begin{abstract}
  We explore an intriguing alternative for a fast and high-fidelity generation of steady-state entanglement.
  By exponentially enhancing the atom-cavity interaction, we obtain an \textit{exponentially-enhanced effective cooperativity} of the
  system, which results in a high fidelity of the state generation. Meanwhile, we modulate the amplitudes of the driving fields
  to accelerate the population transfer to a target state, e.g., a Bell state. An \textit{exponentially shortened stabilization time} is thus predicted.
  Specifically, when the cooperativity of the system is $C=30$, the fidelity of the acceleration scheme reaches $98.5\%$,
  and the stabilization time is about \textit{ten times shorter} than that without acceleration.
  Moreover, we find from the numerical simulation that the acceleration scheme is robust against systematic and stochastic (amplitude-noise) errors.
\end{abstract}


\maketitle

\section{introduction}
\begin{figure}
	\centering
	\scalebox{0.33}{\includegraphics{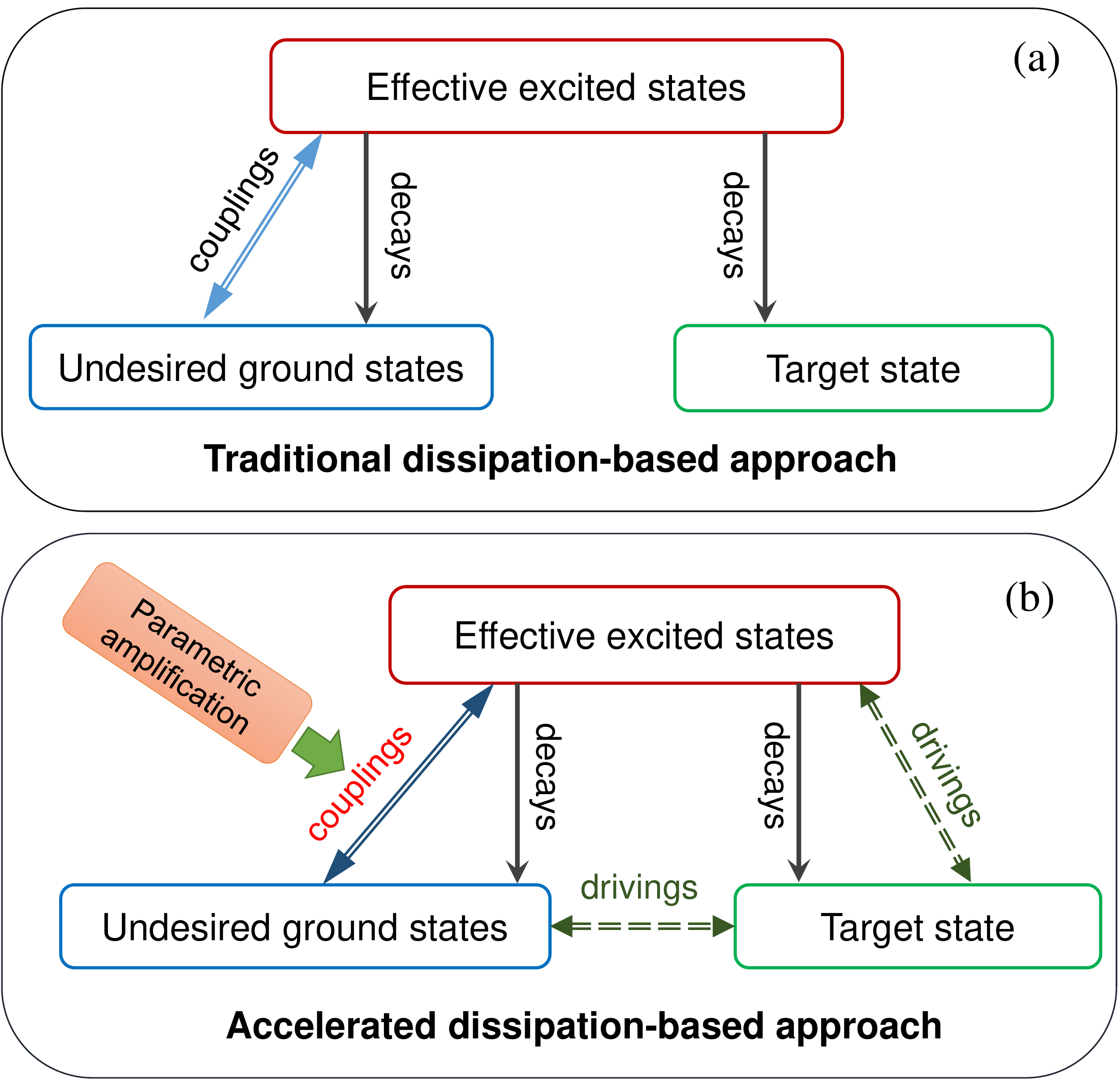}}
	\caption{Schematic of the evolution of a dissipative quantum system in the effective subspace.
		(a) In a traditional dissipation-based (TDB) approach, the system is well engineered, such that the target state is a unique steady state.
		(b) In our accelerated dissipation-based (ADB) approach, we use parametric amplification to improve the effective couplings.
		The (green-dashed arrowed line) additional drivings are induced by pulse modulation, and they are well designed to rapidly increase the population of the target state.
	}
	\label{fig00}
\end{figure}

Quantum entanglement is one of the most striking features of
quantum mechanics, and entangled states of matter
are now widely used for fundamental tests of quantum theory
and applications in quantum information science \cite{Prl852392}. Numerous
schemes have been proposed to faithfully
and controllably generate quantum entangled states \cite{Ajp581131,Pra62062314,Prl802245,Pra63022301} based on either
unitary dynamical evolution \cite{Prl753788,Prl89187903,Prl851762,Prl90027903,Prl96010503,Oe191207,
Oe203176,Pra76062311,Pra89012326,Pra91012325,Pra98062327} or dissipative quantum
dynamical processes
\cite{breuer2002theory,Prl91097905,Njp11083008,Nat4531008,Jpa41065201,Np5633,Prl107120502,Prl110120402,Pra96012315,Prl111033606,Pra85032111,
Nat504419,Prl117040501,Njp14053022,Pra84022316,Pra83042329,Pra82054103,Prl89277901,
Prl117210502,Pra84064302,Nat504415,Prl111033607,Prl106090502,Pra96012328,Npto10303,Prl115200502,Prl107080503,Pra85042306,Sr512753,Pra85062323}.
For convenience, we call the last of these as ``dissipative dynamics'' hereafter.
Dissipative dynamics,
where the dissipation is assumed to be a resource rather than a negative
effect, has recently attracted much interest in quantum computation and entanglement engineering.
The basic idea of a traditional dissipation-based (TDB) approach is shown in Fig.~\ref{fig00}~(a).
Generally, schemes based on dissipative dynamics are robust against parameter fluctuations, can obtain
high-fidelity entanglement with arbitrary initial states, and do
not need accurate control of the evolution time.

In a TDB approach, the key point for entanglement generation is to produce a dissipative
system such that the target state is a unique steady state, regardless of the initial state \cite{Pra85032111,Prl106090502}.
This means that the target state is dropped out of the unitary evolution in the effective subspace. The only way to
transfer population to the target state is via an uncontrollable and slow dissipation process, and the
time required for the entanglement generation is inversely proportional to the decay rates.
Usually, high speed and high fidelity cannot coexist in a TDB approach
because high fidelity $F$ requires high cooperativity $C$, according to $(1-F)\propto 1/\sqrt{C}$ (the optimal value
of the fidelity) \cite{Prl91097905,Prl106090502}, but a high cooperativity means small
decay rates. In addition, for most optical systems, it is usually hard to
achieve a cooperativity of $C$ larger than $100$ \cite{Nt15s556}.
In optical systems, the fast and high-fidelity generation of entangled states in the presence of dissipation is still a challenge in optical systems.

In view of this, we are encouraged to propose a general approach for this problem.
The basic idea of our accelerated dissipation-based (ADB) approach is shown in Fig.~\ref{fig00}~(b).
The parametric amplification based on a squeezed-vacuum field \cite{Prl120093601,Prl120093602}
is used to increase the cooperativity $C$, and as a result to improve the fidelity $F$.
The couplings connecting the undesired ground (UDG) states and the effective excited state are
increased by parametric amplification, but the decays remain unchanged, producing an enhanced cooperativity.
Also, the exponential increase of atom-cavity coupling allows us to choose relatively strong driving fields to shorten the evolution time.
A pulse modulation based on Lyapunov control \cite{Pra80052316,d2007introduction,A4498,Aas331257,A411987,Njp11105034,Pra91032301,Pra88063823,Scl56388} is
used here to induce some additional drivings [the green-dashed arrowed line in Fig.~\ref{fig00}(b)].
The additional drivings are designed to accelerate the population transfer from the UDG states to the target state,
and they gradually vanish when the population of the target state asymptotically reaches $1$.
In this case, the system can be rapidly stabilized into the target state, i.e., the steady entangled state.

With current experimental techniques, it is possible to achieve $C\sim 30$ \cite{Nt15s556,Sci2871447}.
If we consider a relatively good cavity, with cavity decay $\kappa$ smaller than the atomic decay $\gamma$, such as $\kappa\approx0.3\gamma\approx0.1 g$.
Then an evolution time $\geq 1,500/g$ is necessary to achieve a fidelity of $\sim 96\%$ in the TDB approach.
However, by applying our approach, the evolution time is shortened from $1,500/g$ to about $160/g$,
and the final fidelity is improved from $\sim 96\%$ to $\sim 98\%$. Thus, the fast and high-fidelity generation of steady-state entanglement becomes possible.

This paper is organized as follows. In Sec.~II, we show the model and the effective Hamiltonian of the system we consider.
In Sec.~III, we present the ADB approach to realize a fast and high-fidelity generation of steady-state entanglement.
In Sec.~IV, we verify the robustness against parameter errors of the scheme by numerical simulation.
Conclusions are given in Sec.~V.

\begin{figure}
 \centering
 \scalebox{0.23}{\includegraphics{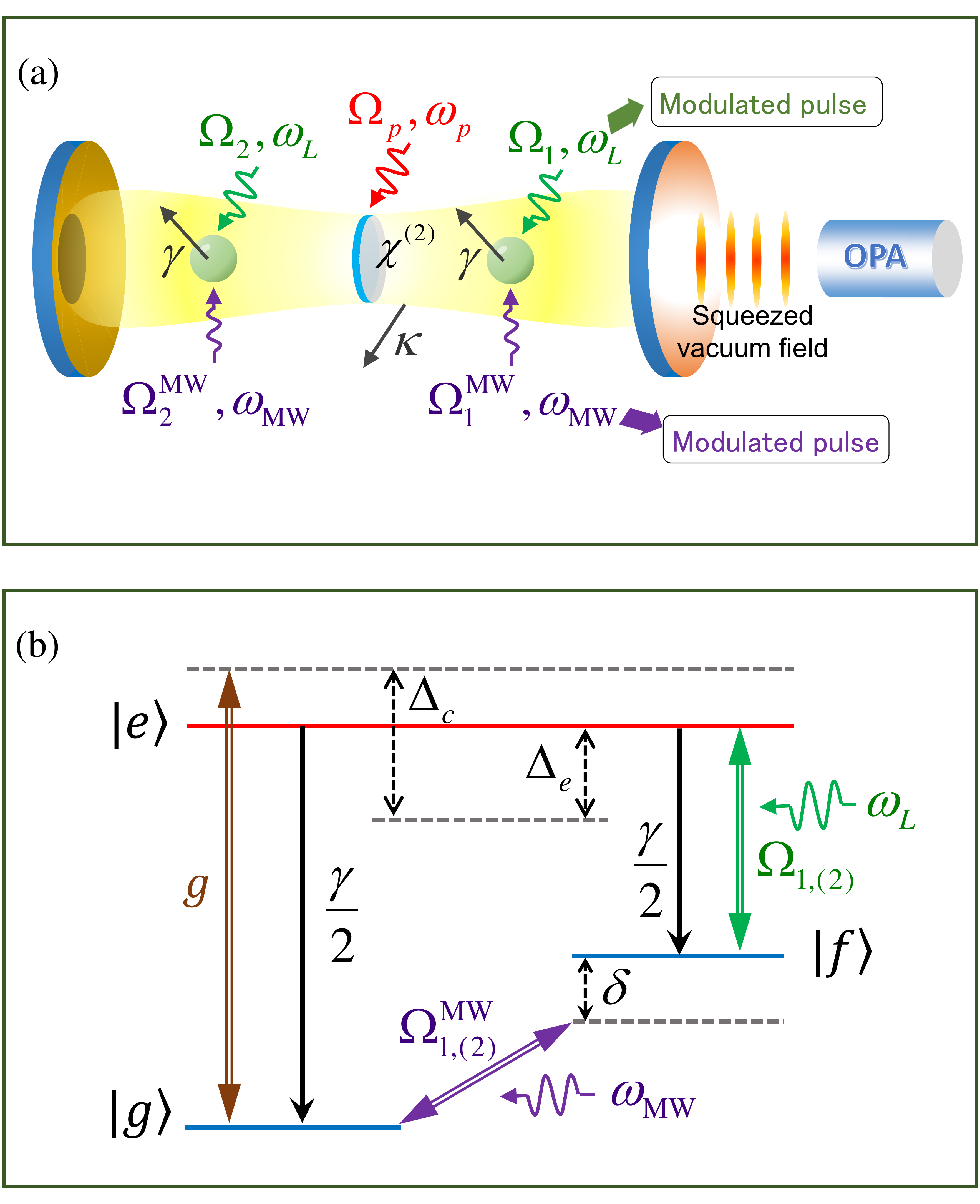}}
 \caption{
         (a) Schematic diagram of a cavity quantum electrodynamics system containing a single-mode cavity of frequency $\omega_{c}$,
             two $\Lambda$ atoms, a $\chi^{(2)}$ nonlinear medium, and an optical parametric amplifier (OPA).
             The OPA is used to generate a squeezed-vacuum reservoir which couples to the cavity.
             Each atom is driven by a laser field and a microwave field with frequencies $\omega_{L}$ and $\omega_{\text{MW}}$, respectively.
             A strong driving field of frequency $\omega_{p}$ is used to pump the nonlinear medium.
             In addition, with pulse modulation, the amplitudes $\Omega_{1}(t)$ and $\Omega_{1}^{\text{MW}}(t)$
             are designed to induce the (green-dashed arrowed line) additional drivings in Fig.~\ref{fig00}.
         (b) For each atom, the ground states $|f\rangle$ and $|g\rangle$ are resonantly driven ({$\omega_{L}=\omega_{e}-\omega_{f}$}) and off-resonantly
             coupled to the excited state $|e\rangle$ with Rabi frequency $\Omega_{j}(t)$ and coupling $g$, respectively.
             The detunings are $\Delta_{e}=\omega_{e}-\omega_{g}-\omega_{p}/2$, $\Delta_{c}=\omega_{c}-\omega_{p}/2$, and $\delta=\omega_{f}-\omega_{g}-\omega_{\text{MW}}$, respectively,
             where $\omega_{z}$ is the frequency associated with level $|z\rangle$ ($z=g,f,e$).
             For convenience, we assume that the spontaneous emission rates are the same for decaying to the $|g\rangle$ and to the $|f\rangle$ states (i.e., $\gamma_{f}=\gamma_{g}=\gamma/2$).
         }
 \label{fig0}
\end{figure}

\section{Model}
As shown in Fig.~\ref{fig0}(a),
we consider a quantum system with two $\Lambda$ atoms trapped in a single-mode cavity.
The level structure of each atom is shown in Fig.~\ref{fig0}(b). Note that the pulse modulation is only applied to one of the atoms.
The Hamiltonian determining the unitary dynamics of the system, via the rotating wave approximation in a proper observation frame, reads ($\hbar=1$)
\begin{align}\label{eq1-01}
  H_{0}=&\sum_{j=1,2}{\Delta_{e}|e\rangle_{j}\langle e|}+H_{\text{AC}}+H_{\text{NL}}+V+H_{g}, \cr
  H_{\text{AC}}=&\sum_{j=1,2}g|e\rangle_{j}\langle g|a+\text{H.c.} \cr
  H_{\text{NL}}=&\Delta_{c}a^{\dag}a+\Omega_{p}(e^{i\theta_{p}}a^{2}+\text{H.c.}).
\end{align}
Here, $V=\sum_{j=1,2}{}\Omega_{j}(t)e^{-i\Delta_{e}t}|e\rangle_{j}\langle f|+\text{H.c.}$
describes the interaction of a classical laser drive with the atoms,
$H_{g}=\sum_{j=1,2}{\Omega_{j}^{\text{MW}}(t)e^{i\delta t}|f\rangle_{j}\langle g|+\text{H.c.}}$ describes the interaction between the ground states.
For brevity, we omit the explicit time dependence of the Hamiltonians $H_{0}$, $H_{g}$ and $V$.

By introducing the Bogoliubov squeezing transformation $a_{\text{sc}}=\cosh(r_{p})a+e^{-i\theta_{p}}\sinh(r_{p})a^{\dag}$,
we diagonalize $H_{\text{NL}}$ as $H_{\text{NL}}=\omega_{\text{sc}}a_{\text{sc}}^{\dag}a_{\text{sc}}$, where
\begin{align}\label{eq1-011}
  r_{p}=\frac{1}{4}\ln{\frac{1+\alpha}{1-\alpha}},
\end{align}
{is the squeezing parameter of the squeezed-cavity mode}, and
$\omega_{\text{sc}}=\Delta_{c}\sqrt{1-\alpha^2}$ is the squeezed-cavity frequency ($\alpha=2\Omega_{p}/\Delta_{c}$).
In this case,
when $g\sinh(r_{p})\ll(\omega_{\text{sc}}+\Delta_{e})$ and $\Delta_{e}=\omega_{\text{sc}}$,
we obtain the exponentially-enhanced atom-cavity coupling (see the Appendix for details)
\begin{align}
  g_{\text{sc}}=g\cosh(r_{p}),
\end{align}
and the atom-squeezed-cavity interaction Hamiltonian
\begin{align}\label{eq1-02}
  H'_{\text{AC}}=g_{\text{sc}}\sum_{j=1,2}{a_{\text{sc}}|e\rangle_{j}\langle g|+\text{H.c.}}.
\end{align}

\begin{figure}
	\centering
	\scalebox{0.28}{\includegraphics{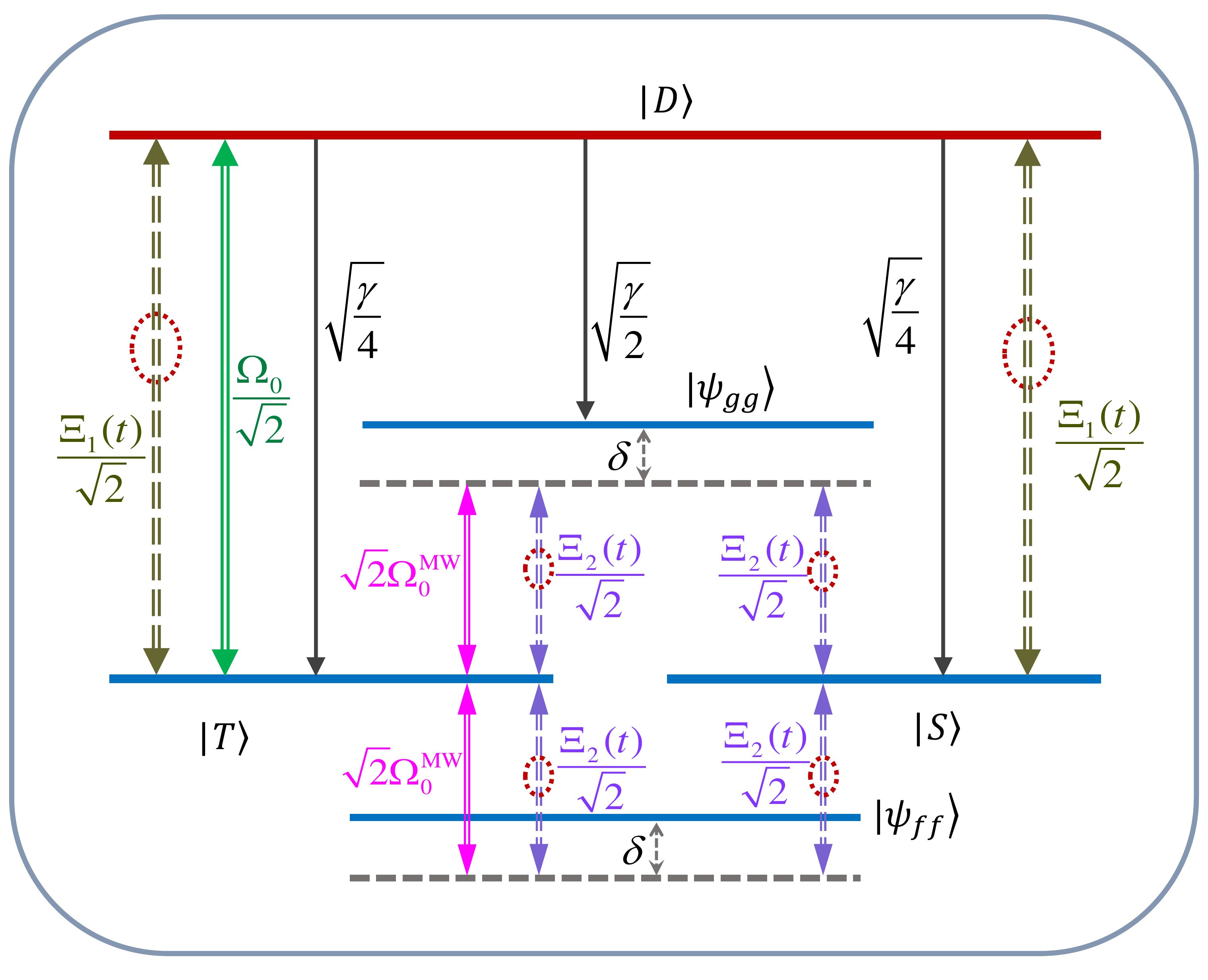}}
	\caption{
		The effective transitions for the two-atom system when $\Omega_{j}(t),\Omega_{j}^{\rm{MW}}(t)\ll g_{\rm{sc}}$.
		With the effective driving fields and decays,
		ultimately, the system will be stabilized into the state $|S\rangle$.
		Here, the red-dotted ellipses represent the control fields induced by the pulse modulation.
	}
	\label{figb1}
\end{figure}

We squeeze the cavity mode to exponentially enhance the atom-cavity coupling, as described above.
{This can introduce additional noise into the cavity \cite{Prl120093601,Prl120093602}.
This additional noise can be understood as an effective thermal noise and an effective two-photon correlation.
To circumvent such undesired noise, we introduce a squeezed-vacuum field by an optical parametric amplifier [see Fig.~\ref{fig0}(b)], with
a squeezing parameter $r_{e}$ and a reference phase $\theta_{e}$, to drive the cavity.}
When choosing $r_{e}=r_{p}$ and $\theta_{e}+\theta_{p}=\pm \left(2n+1\right)\pi$ ($n=0,1,2,\ldots$),
we can completely eliminate this additional noise, as detailed in the Appendix.
{In this case, the squeezed-cavity mode is equivalently coupled to a thermal reservoir and the
additional noise is completely removed.}
Thus, we can use a standard Lindblad operator to describe the
squeezed-cavity decay, i.e., $L_{\text{sc}}=\sqrt{\kappa}~a_{\text{sc}}$. The system in this case can be modeled by a master equation in the Lindblad form \cite{Rmp3247,Cmp48119}:
\begin{align}\label{eq1-03}
  \dot{\rho}=&i[\rho,H_{0}]+\mathcal{L}(\rho), \cr
  \mathcal{L}(\rho)=&\sum_{k}{L_{k}\rho L_{k}^{\dag}-\frac{1}{2}(L_{k}^{\dag}L_{k}\rho+\rho L_{k}^{\dag}L_{k})},
\end{align}
where 
$L_{k}$'s are the lindblad operators describing a cavity decay $L_{\text{sc}}=\sqrt{\kappa}a_{\text{sc}}$,
and four spontaneous emissions $L_{j,z'}=\sqrt{\gamma/2}|z'\rangle_{j}\langle e|$ ($j=1,2$, $z'=g,f$).
Consequently, increasing $r_{p}$ enables an exponential enhancement in the cooperativity,
\begin{align}\label{eq1-003}
  \frac{C_{\text{sc}}}{C}=\cosh^{2}{(r_{p})}.
\end{align}
When $r_{p}>1$, we have
\begin{align}\label{eq}
  \frac{C_{\rm{sc}}}{C}\simeq  \frac{\exp{(2r_{p})}}{4}.
\end{align}

{Assuming that the squeezed cavity is initially in the squeezed vacuum
state $|0\rangle_{\rm{sc}}$ and the atoms are initially in their ground states,}
in the limit of $\Omega_{j}(t),\Omega_{j}^{\text{MW}}(t)\ll g_{\rm{sc}}$
\cite{Jmp18756,Pra412295,Prl89080401,Jpa41493001,Pst2149,Prl744763,Pra77062339},
the evolution of the system is confined to an effective evolution subspace  
spanned by $|\psi_{gg}\rangle=|gg\rangle|0\rangle_{\rm{sc}}$, $|\psi_{ff}\rangle=|ff\rangle|0\rangle_{\rm{sc}}$,
\begin{align}\label{eq1-04}
   |T\rangle=&(|fg\rangle+|gf\rangle)|0\rangle_{\rm{sc}}/\sqrt{2}, \cr
   |S\rangle=&(|fg\rangle-|gf\rangle)|0\rangle_{\rm{sc}}/\sqrt{2}, \cr
   |D\rangle=&(|eg\rangle-|ge\rangle)|0\rangle_{\rm{sc}}/\sqrt{2}.
\end{align}
Meanwhile, the
decay process in this subspace can be described by three effective Lindblad operators \cite{Prl106090502,Pra96012328,arxiv180907324}
\begin{align}\label{eq1-005}
  \tilde{L}_{G}=\sqrt{\frac{\gamma}{2}}|\psi_{gg}\rangle\langle D|, \ \
  \tilde{L}_{T,(S)}=\sqrt{\frac{\gamma}{4}}|T(S)\rangle\langle D|.
\end{align}
Here, the cavity mode has been adiabatically eliminated in the limit of $\Omega_{j}(t),\Omega_{j}^{\rm{MW}}(t)\ll g_{\rm{sc}}$.
Note that, here, although in the laboratory frame the squeeze-cavity mode contains a large number of photons,
the cavity degree of freedom is adiabatically eliminated in our proposal, resulting in a squeezed-cavity mode mediated coupling between atoms.
Thus, our proposal can be potentially extended to implementations of various intracavity quantum operations

\section{Fast and high-fidelity entanglement generation}
\begin{figure}
	\centering
	\scalebox{0.27}{\includegraphics{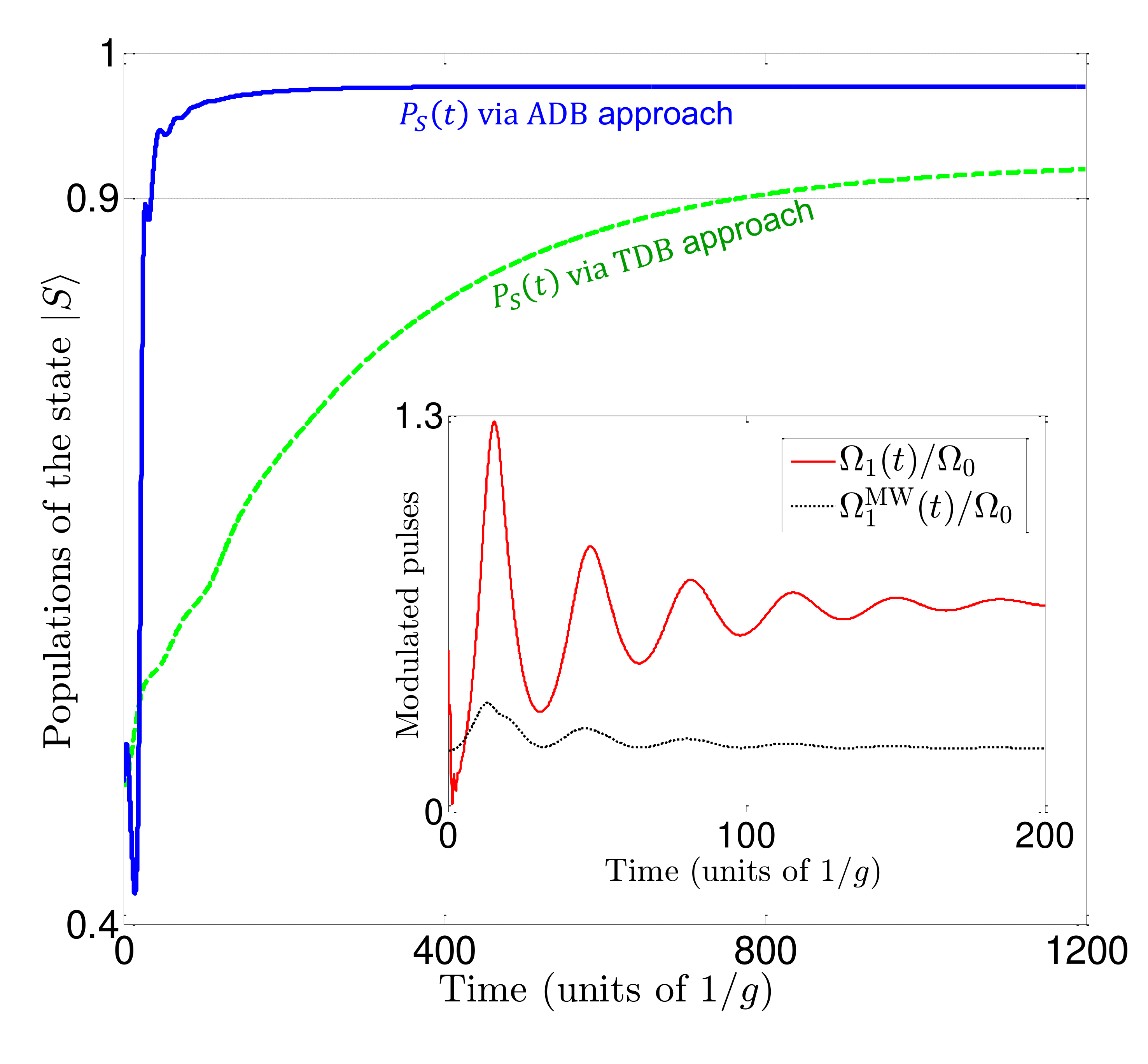}}
	\caption{
		The population increases of the target state $|S\rangle$ in the ADB approach (blue-solid curve) and in the TDB approach (green-dashed curve).
		The inset shows the modulated pulses for the ADB approach.
		The parameters used here are $K_{1}=g$, $K_{2}=0.15g$, $r_{p}=2$, $\Omega_{0}=0.1g_{\text{sc}}$,
		$\Omega^{\rm{MW}}_{0}=0.2\Omega_{0}$, $\delta=0.4\Omega_{0}$, $\kappa=0.3\gamma$, and $C=g^{2}/(\kappa\gamma)=30$.
	}
	\label{fig01}
\end{figure}

We assume that the Rabi frequencies $\Omega_{j}(t)$ and $\Omega_{j}^{\text{MW}}(t)$ are
\begin{align}\label{eq1-a121}
  \Omega_{1}(t)=&\Omega_{0}/\sqrt{2}+\Xi_{1}(t),\cr
  \Omega_{1}^{\text{MW}}(t)=&\Omega^{\rm{MW}}_{0}/\sqrt{2}+\Xi_{2}(t), \cr
  \Omega_{2}(t)=&e^{i\pi}\Omega_{0}/\sqrt{2},\cr
  \Omega_{2}^{\text{MW}}(t)=&\Omega_{0}^{\rm{MW}}/\sqrt{2},
\end{align}
where $\Omega_{0}$ and $\Omega^{\rm{MW}}_{0}$ are constants,
$\Xi_{1,(2)}(t)$ is the control function of the pulse modulation.
We show the effective transitions of the system in Fig.~\ref{figb1}.
When the pulse modulation is implemented, the evolution process of
the system can be described as follows:
\begin{basedescript}{\desclabelstyle{\pushlabel}\desclabelwidth{1em}}
	\item[$\bullet$]  The microwave fields $\Omega_{j}^{\rm{MW}}(t)$ directly drive the population transfer between the ground states,
	                  so that the populations cannot be stored in the UDG states.
	\item[$\bullet$] Once the population is transferred to the state $|T\rangle$, the modulated driving field $[\Omega_{0}+\Xi_{1}(t)]/\sqrt{2}$
	                 will drive the transition $|T\rangle\rightarrow|D\rangle$.
	\item[$\bullet$] Then, the population in state $|D\rangle$ will be transferred to $|S\rangle$ via the decay $\tilde{L}_{S}$ and the driving $\Xi_{1}(t)$.
\end{basedescript}
In this case, by suitably adjusting the control functions $\Xi_{j}(t)$, we can achieve the target state $|S\rangle$ in a very short time.
{The shapes of the modulated pulses,
as shown in the inset of Fig. \ref{fig01}, are shown to be smooth time-dependent curves that are realizable experimentally.
For example, experiments used electro-optic modulators to implement such control fields \cite{Nc712479}.}

Motivated by Lyapunov control theory \cite{d2007introduction,A4498,Aas331257,A411987,Njp11105034,Pra91032301,Pra88063823}, we define the speed of
the population increase for a state as the time derivative of its population, i.e.,
\begin{align}\label{eq1-b1}
  \mathcal{V}_{x}(t)=\dot{P}_{x}(t)=\text{Tr}(\dot{\rho}\rho_{x}),
\end{align}
where $\rho_{x}=|x\rangle\langle x|$. The rates of the population increase for the ground states are
\begin{align}\label{eq1-b2}
  \mathcal{V}_{S}(t)=&\frac{\gamma}{4}\langle D|\rho| D\rangle-i\text{Tr}\left\{[\rho_{S},H_{\rm{mod}}]\rho\right\},      \cr
  \mathcal{V}_{T}(t)=&\frac{\gamma}{4}\langle D|\rho|D\rangle-i\text{Tr}\left\{[\rho_{T},(H_{\rm{s}}+H_{\rm{mod}})]\rho\right\}, \cr
  \mathcal{V}_{g}(t)=&\frac{\gamma}{2}\langle D|\rho|D\rangle-i\text{Tr}\left\{[\rho_{gg},(H_{\rm{s}}+H_{\rm{mod}})]\rho\right\}, \cr
  \mathcal{V}_{f}(t)=&-i\text{Tr}\left\{[\rho_{ff},(H_{\rm{s}}+H_{\rm{mod}})]\rho\right\},
\end{align}
respectively.
Here, the Hamiltonian
\begin{align}\label{eq1-b23}
  H_{\rm{s}}=&\frac{1}{\sqrt{2}}[\Omega_{0}| D\rangle+2\Omega^{\rm{MW}}_{0}(e^{i\delta t}|\psi_{gg}\rangle+ e^{-i\delta t}|\psi_{ff}\rangle)]\langle T|\cr
               &+\text{H.c.},
\end{align}
is the effective Hamiltonian of the system when $\Xi_{1,(2)}(t)=0$.
{On account of $H_{\rm{s}}|S\rangle=0$, $\tilde{L}_{k}|S\rangle=0$, and $\tilde{L}_{k}|S\rangle\neq0$ ($k=L,S,G$),}
there is a unique steady state, i.e., the target entangled state $|S\rangle$, for the system when $\Xi_{1,(2)}(t)=0$.
The Hamiltonian
\begin{align}\label{eq1-b24}
  H_{\rm{mod}}=&\Xi_{1}(t)H_{1}+\Xi_{2}(t)H_{2},\cr
  H_{1}=&|f\rangle_{1}\langle e|+|e\rangle_{1}\langle f|,\cr
  H_{2}=&|f\rangle_{1}\langle g|+|g\rangle_{1}\langle f|,
\end{align}
describes the interaction induced by the pulse modulation.
Substituting $H_{1}$ and $H_{2}$ into Eq.~(\ref{eq1-b2}), we find the following:
\begin{basedescript}{\desclabelstyle{\pushlabel}\desclabelwidth{1em}}
\item[(i)] $[\rho_{S},H_{m}]\neq 0$ and $[\rho_{T},H_{m}]\neq 0$ ($m=1,2$);
    both the control functions $\Xi_{m}(t)$ can adjust the speeds $\mathcal{V}_{S}(t)$ and $\mathcal{V}_{T}(t)$;
\item[(ii)]$[\rho_{gg},H_{1}]= 0$, $[\rho_{gg},H_{2}]\neq 0$, $[\rho_{ff},H_{1}]= 0$, and $[\rho_{ff},H_{2}]\neq 0$; the
    control function $\Xi_{2}(t)$ can adjust the speeds $\mathcal{V}_{gg}(t)$ and $\mathcal{V}_{ff}(t)$, while $\Xi_{1}(t)$ cannot.
\end{basedescript}
These two points indicate that, it is hard to design $\Xi_{j}(t)$ to control one of the speeds in Eq.~(\ref{eq1-b2}) without influencing the others.

\begin{figure}
	\centering
	\scalebox{0.285}{\includegraphics{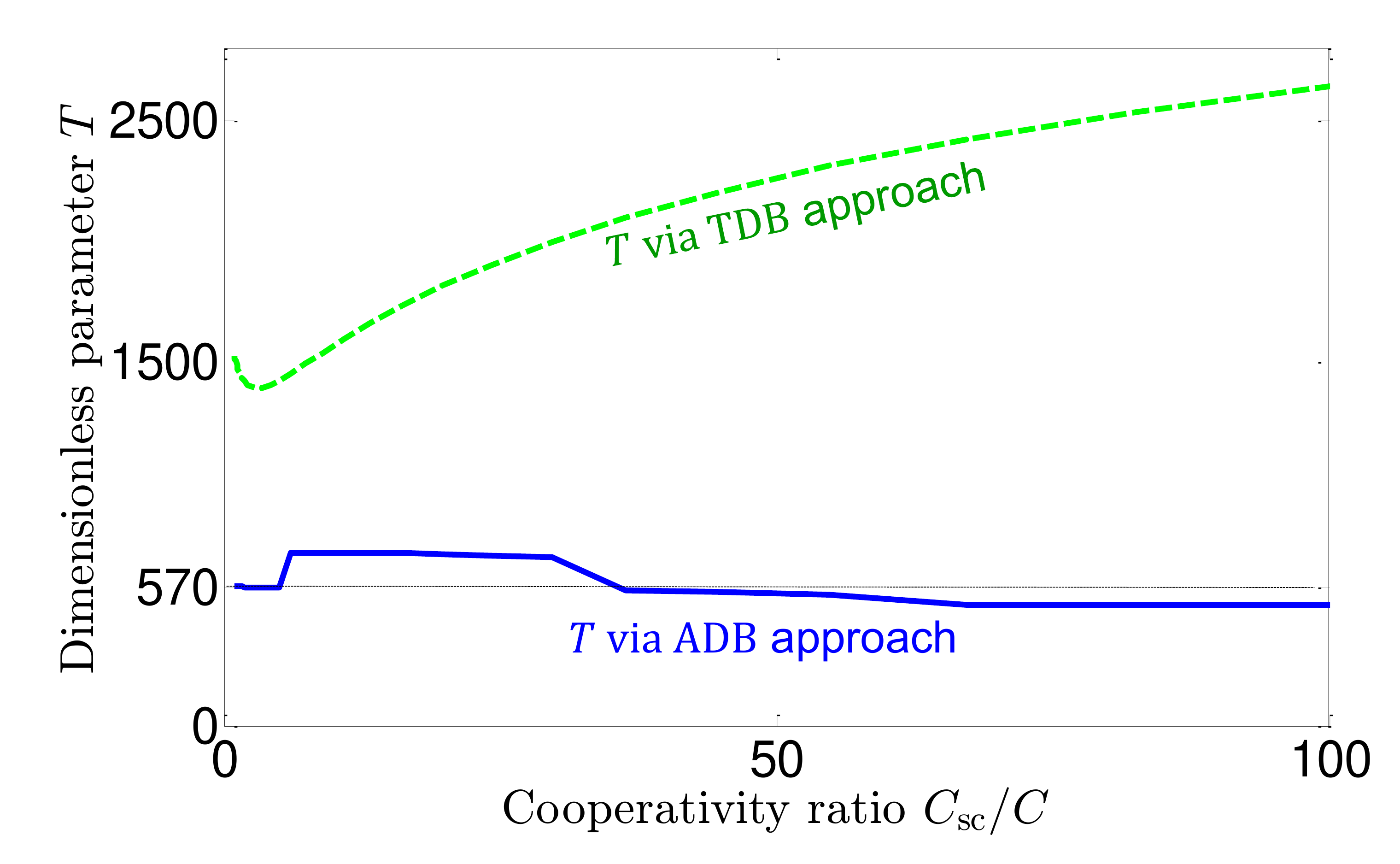}}
	\caption{
		The dimensionless parameter $T=(t_{S}\cdot g_{\rm{sc}})$ versus the cooperativity ratio $C_{\rm{sc}}/C=\cosh^{2}(r_{p})$ using different dissipation-based approaches.
		Here, except $r_{p}$, the parameters are the same as the set in Fig.~\ref{fig01}.
	}
	\label{fig05}
\end{figure}

In this case, a simple choice is designing $\Xi_{j}(t)$ to only control the speed $\mathcal{V}_{S}(t)$,
i.e., the control functions $\Xi_{j}(t)$ are designed as
\begin{align}\label{eq1-6a}
  \Xi_{j}(t)=-iK_{j}\text{Tr}\left\{[\rho_{S},H_{j}]\rho\right\},
\end{align}
where $K_{j}>0$. Thus, the second term in $\mathcal{V}_{S}(t)$
is positive, $-i\text{Tr}\left\{[\rho_{S},H_{\rm{mod}}]\rho\right\}\geq0$, and the
speed $\mathcal{V}_{S}(t)$ is improved.
However, since the driving $\Xi_{2}(t)$ cannot directly induce an entanglement,
designing $\Xi_{2}(t)$ according to the target state $|S\rangle$ is not the best choice for our goal \cite{Pra97032328}.
In view of this, we have to seek for a new way to design the control function $\Xi_{2}(t)$.

\renewcommand\arraystretch{1.5}
\begin{table*}[t]
	\centering
	\caption{Comparison between schemes with and without pulse modulation and parametric amplification.
		Note the very significant decrease in the stabilization time $t_{S}$, and the increase in the fidelity $F$.~\\ }
	\label{tab1}
	\begin{tabular}{l|p{2cm}<{\centering}p{2cm}<{\centering}p{2cm}<{\centering}p{2cm}<{\centering}}
		\hline
		\hline
		Dissipation-based schemes     & Squeezing parameter \ \ \ \ \ \  $r_{p}$ & Cooperativity rate \ \ \ \ \ \  $C_{\rm{sc}}/C$ & Stabilization time \ \ \ \ \ \ \ \ \ \ \ \ \  $t_{S}$ & Fidelity \ \ \ \ \ \ \ \ \ \ \ \ \ \ \ \  $F$ \\
		\hline
		Via traditional method       & 0 & 1  & $\sim1500/g$ & $\sim96\%$  \\
		Via pulse modulation          & 0 &  1 & $\sim 570/g$ & $\sim95\%$ \\
		Via parametric amplification  & 2 & 14 & $\sim400/g$  & $\sim98.7\%$ \\
		Via our acceleration method   & 2 & 14 & $\sim 160/g$ & $\sim98.6\%$ \\
		\hline
		\hline
	\end{tabular}
\end{table*}

It is worth noting that the decay causes a relatively fast population increase in the UDG state $|\psi_{gg}\rangle$
according to Eq.~(\ref{eq1-b2}), and more population will
be decayed to the state $|\psi_{gg}\rangle$ than $|S\rangle$ after a certain time evolution.
A slow evolution is inevitable to totally transfer the population from the state $|\psi_{gg}\rangle$ to the target state $|S\rangle$.
By considering this, the control function $\Xi_{2}(t)$ can be chosen as
\begin{align}\label{eq2-a1}
  \Xi_{2}(t)=iK_{2}\text{Tr}\left\{{[\rho_{gg},H_{2}]\rho}\right\},
\end{align}
which is designed to decrease the population of the state $|\psi_{gg}\rangle$ by adding
a negative term  to the speed $\mathcal{V}_{g}(t)$.
Then, the evolution speeds $\mathcal{V}_{S}(t)$ and $\mathcal{V}_{g}(t)$ read
\begin{align}\label{eq2-a2}
  \mathcal{V}_{S}(t)=&\frac{\gamma}{4}\langle D|\rho|D\rangle+\frac{|\Xi_{1}(t)|^{2}}{K_{1}}-i\Xi_{2}(t)\text{Tr}\left\{[\rho_{S},H_{2}]\rho\right\}, \cr
  \mathcal{V}_{g}(t)=&\frac{\gamma}{2}\langle D|\rho|D\rangle-\frac{|\Xi_{2}(t)|^{2}}{K_{2}}-i\text{Tr}\left\{[\rho_{gg},H_{\rm{s}}]\rho\right\},
\end{align}
respectively. Due to $[\rho_{S},H_{2}]\neq 0$, the last term in $\mathcal{V}_{S}(t)$ may have a negative effect on the evolution speed, but
we can adjust the parameters $K_{1}$ and $K_{2}$ to minimize this negative effect.

\begin{figure}
	\centering
	\scalebox{0.28}{\includegraphics{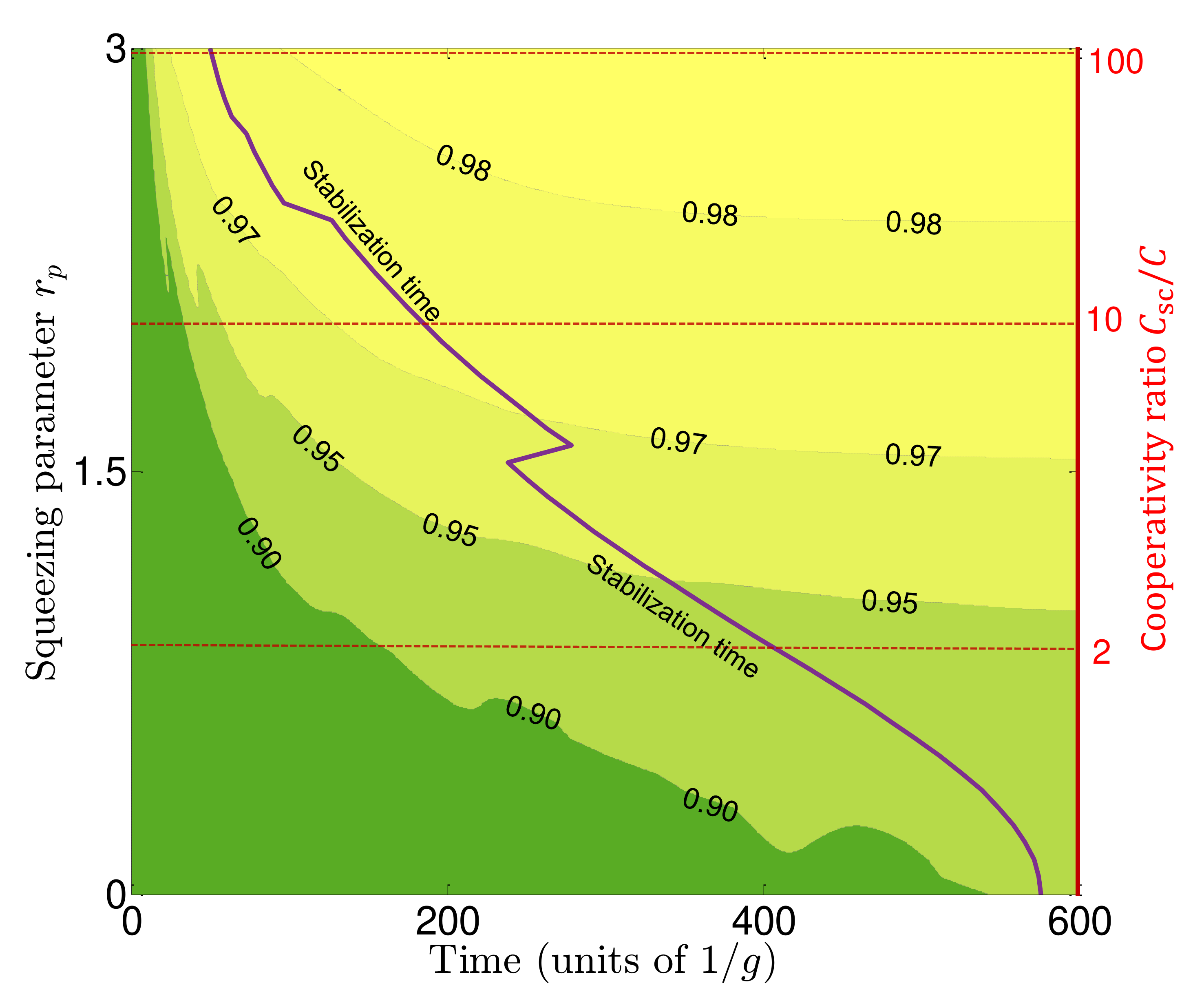}}
	\caption{
		Contour plot of the population $P_{S}(t)$ versus time and squeezing parameter $r_{p}$ in the ADB approach.
		The red vertical axis (on the right) denotes the cooperativity ratio $C_{\text{sc}}/C=\cosh^{2}(r_{p})$.
		The purple-solid curve represents the time $t_{S}$ when the system becomes stable.
		Here, $r_{p}$ is an independent variable (in the left vertical axis), and other parameters are the same as in Fig.~\ref{fig01}.
	}
	\label{fig02}
\end{figure}

A comparison between the TDB method and the ADB approach is shown in Fig.~\ref{fig01}.
It takes a very short time (about $110/g$) in the ADB approach to generate the target state with population $\sim 90\%$,
while it takes a much longer time (about $780/g$) in the TDB scheme.
In the ADB approach, when $t\geq t_{S}= 160/g$, the system gradually becomes stable. The time $t_{S}$ is called
the ``stabilization time,'' and it describes the time when the system becomes stable.
Here, the stabilization is determined according to
$\mathcal{V}_{S}(t_{S})\rightarrow 0$ and $\dot{\mathcal{V}}_{S}(t_{S})\rightarrow 0$.
Specifically, in this paper, we assume that when $\mathcal{V}_{S}(t_{S})\leq10^{-5}g$ and $\dot{\mathcal{V}_{S}}(t_{S})\leq10^{-6}g^2$,
the system is stable.

For brevity, we define a dimensionless parameter
\begin{align}
  T=t_{S}\cdot g_{\rm{sc}},
\end{align}
representing a measurement scale of the stabilization time in the following analysis.
As shown in Fig.~\ref{fig05}, the dimensionless parameter $T$ in the TDB scheme increases
when the amplified cooperativity $C_{\rm{sc}}$ increases,
for example, $T\approx1,500$ when $C_{\rm{sc}}=30$, and $T\approx2,500$ when $C_{\rm{sc}}=3,000$.
However, we find the relationship between
the stabilization time $t_{S}$ and the squeezing parameter $r_{p}$
(see the purple-solid curve in Fig.~\ref{fig02}) in our ADB approach is
\begin{align}\label{eq1-7}
  t_{S}\approx \frac{570}{g\cosh{(r_{p})}}=\frac{570}{g_{\text{sc}}},
\end{align}
which means that ${T}\approx 570$ (see the blue-solid curve in Fig.~\ref{fig05}) is independent of the amplified cooperativity $C_{\rm{sc}}$.
This is an important result of this paper. It predicts
an \textit{exponentially-shortened stabilization time} $t_{S}$  when $r_{p}>1$.
Moreover, the comparison between the TDB approach and the ADB approach in Fig.~\ref{fig05} indicates that the
pulse modulation works better in accelerating the evolution when the amplified cooperativity $C_{\rm{sc}}$ is larger.
This means that \textit{the pulse modulation and the parametric amplification supplement each other} in the ADB approach to realize
a fast and high-fidelity generation of steady-state entanglement. This result is also shown in Table~\ref{tab1}, which shows the
comparison between methods with and without pulse modulation and parametric amplification.
The improvements in the two bottom rows are very significant.
In Table~\ref{tab1}, the fidelity $F$ is defined as
\begin{align}\label{eq3-8}
  F=\sqrt{\langle S|\rho(t_{f})|S\rangle},
\end{align}
where $t_{f}$ denotes the final time. For convenience, we set $t_{f}\equiv t_{S}$ in this paper.

The final population of the state $|S\rangle$ increases when
the squeezing parameter $r_{p}$ becomes larger (see Fig.~\ref{fig02}).
When $r_{p}=3$, the cooperativity is amplified to $C_{\text{sc}}\approx100C=3,000$,
and the population of the target state $|S\rangle$ can reach $P_{S}(t_{S})\geq97.5\%$.
The stabilization time $t_{S}=5/g$ is 6 times shorter than
that in Ref. \cite{Prl120093601} by only using parametric amplification.
Generally, the fidelity of a dissipation-based scheme is higher when $r_{p}$ is larger.
However, a large squeezing parameter $r_{p}$ corresponds to an extremely strong $\Omega_{p}$.
For example, when $r_{p}=3$, the driving field $\Omega_{p}$ reaches $\Omega_{p}> 10^{4}g$, which
may cause problems in some experiments.
It is better to choose the squeezing parameter $r_{p}\leq 2$ corresponding to $\Omega_{p}\leq 10^{2}g$.
When $r_{p}=2$, the stabilization time becomes $t_{S}\approx 160/g$, which is almost 10 times shorter than that obtained via traditional method as shown in Table~\ref{tab1}.

\section{Robustness against parameter errors}
Influenced by the environment, there are
usually two kinds of parameter errors, which should be considered
in realizing this approach: systematic error and stochastic error.
It is usually hard to avoid errors; for example, the atoms might not be
ideally placed. Thus, the various atoms may be subject to slightly
different fields, which causes a systematic error.
In this case, the actual Hamiltonian should be corrected as $H_{\rm{n}}=H_{0}+\lambda H_{e1}$, where the subscript
``n' represents the ``noise,'' $\lambda$ is the amplitude of the systematic noise, and $H_{e1}$ is a perturbed Hamiltonian.

\begin{figure}
	\centering
	\scalebox{0.28}{\includegraphics{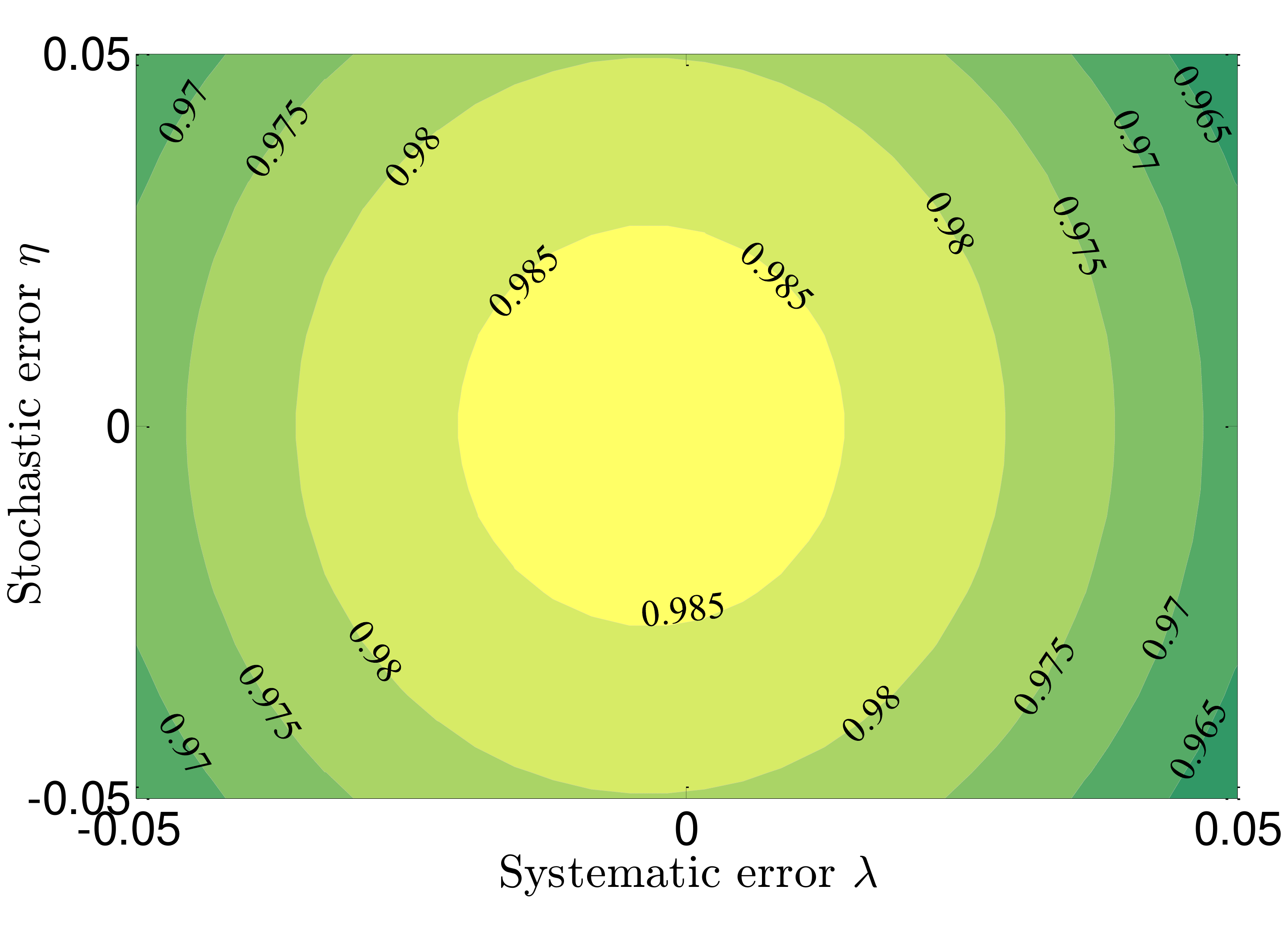}}
	\caption{
		Contour plot of the fidelity $F$
		versus systematic error $\lambda$ and stochastic error $\eta$.
		Here, the final time is chosen as $t_{f}=160/g$, and other parameters are the same as in Fig.~\ref{fig01}.
	}
	\label{fig03}
\end{figure}

When the stochastic error is considered, the actual Hamiltonian becomes $H_{\rm{n}}=H_{0}+\lambda H_{e1}+\eta H_{e2}\xi(t)$,
where $\xi{(t)}=\frac{\partial}{\partial t}W_{t}$ is the
time derivative of the Brownian motion $W_{t}$, $\eta$ is the amplitude of the stochastic noise, and $H_{e2}$ is also a perturbed Hamiltonian.
Since the noise should have zero mean and the noise
at different times should be uncorrelated, we have $\langle\xi(t)\rangle=0$ and $\langle\xi(t)\xi(t')\rangle=\delta(t-t')$.
Then, the master equation of the system in the presence of noise is
\begin{align}\label{eq2-1}
  \dot{\rho}_{\rm{n}}=&-i[H_{\rm{n}},\rho_{\rm{n}}]+\mathcal{L}\rho_{n} \cr
                     =&-i[H_{0},\rho_{\rm{n}}]+\mathcal{L}\rho_{n}\cr
                      &-i\lambda[H_{e1},\rho_{\rm{n}}]-i\eta[H_{e2},\xi(t)\rho_{\rm{n}}].
\end{align}
By averaging over the noise, Eq.~(\ref{eq2-1}) becomes
\begin{align}\label{eq2-2}
  \dot{\rho}_{\rm{n}}=&-i[H_{0},\rho_{\rm{n}}]+\mathcal{L}\rho_{n}\cr
                      &-i\lambda[H_{e1},\rho_{\rm{n}}]-i\eta[H_{e2},\langle\xi(t)\rho_{\rm{n}}\rangle].
\end{align}
According to Novikov's theorem applied to white noise \cite{Pra96043853,Njp14093040}, we have
\begin{align}\label{eq2-a3}
  \langle\xi(t)\rho_{\rm{n}}\rangle=\frac{1}{2}\left\langle{\frac{\delta \rho_{\rm{n}}}{\delta\xi(s)}}\right\rangle_{s=t}=-\frac{i\lambda}{2}[H_{e2},\rho_{\rm{n}}].
\end{align}

We assume the presence of systematic and stochastic (amplitude-noise) errors
in $\Omega_{1}(t)$, so that $H_{e1}=H_{e2}=\Omega_{1}(t)(|e\rangle_{1}\langle f|+|f\rangle_{1}\langle e|)$.
For the ADB approach, we can choose $t_{f}=160/g$,
which is enough for the population $P_{S}$ to reach $P_{S}\simeq97\%$ when $r_{p}=2$, as shown above.
The systematic error $\lambda$ has a more serious influence on the fidelity than the stochastic error $\eta$, as shown in Fig.~\ref{fig03}.
A systematic error with intensity $\lambda=0.05$ causes a deviation of about $1.5\%$ on the fidelity, while a same-intensity stochastic error only causes
$\sim0.5\%$ deviation. When $\lambda=\eta=0.05$, the fidelity is still higher than $96\%$, which demonstrates that the ADB approach is robust against both systematic and stochastic errors.
Beware that the control functions $\Xi_{1}(t)$ and $\Xi_{2}(t)$ in $H_{0}$ must be given according to the master
equation without the noise terms; otherwise, the numerical simulation result in Fig.~\ref{fig03} is possibly wrong.

\section{Possible implementations}
In a cavity quantum electrodynamics system, as shown in Fig.~\ref{fig0}(a), we consider a
possible experimental implementation with ultracold $^{87}$Rb atoms trapped in a single-mode Fabry-Perot cavity.
The $^{87}$Rb atoms can be used for the $\Lambda$-type qutrits as shown in Fig.~\ref{fig0}(b).
Focusing on the $D_{1}$ line electric-dipole transitions at a wavelength of $795$ nm,
the excited state $|e\rangle$ corresponds to the $F'=2,\ m'_{F}=-2$ hyperfine state of the $5^{2}P_{1/2}$ electronic state, and
the ground states $|f\rangle$ and $|g\rangle$ correspond to the $F=1,\ m_{F}=-1$ and the $F=2,\ m_{F}=-1$ hyperfine states of the $5^{2}S_{1/2}$ electronic ground states, respectively.
The transition $|f\rangle\leftrightarrow|e\rangle$ is coupled by a circularly $\sigma^{-}$-polarized control laser. The
transition $|g\rangle\leftrightarrow|e\rangle$ is coupled by a $\pi$-polarized-cavity mode.
The transition $|f\rangle\leftrightarrow|g\rangle$ is electric-dipole forbidden, as it is the case for hyperfine levels of
alkali atoms. However a magnetic dipole transition may be used
instead, although this may limit the intensity (hundreds of kHz as reported in Refs.~\cite{Fp54702,Pra90053416}).
In this case, according to the experimental parameters \cite{Sci2871447},
the fidelity of the ADB approach in Sec.~III can reach $F\geq 98\%$, and the corresponding stabilization time is $t_{S}\approx 11$ $\mu$s.

\begin{figure}
	\centering
	\scalebox{0.28}{\includegraphics{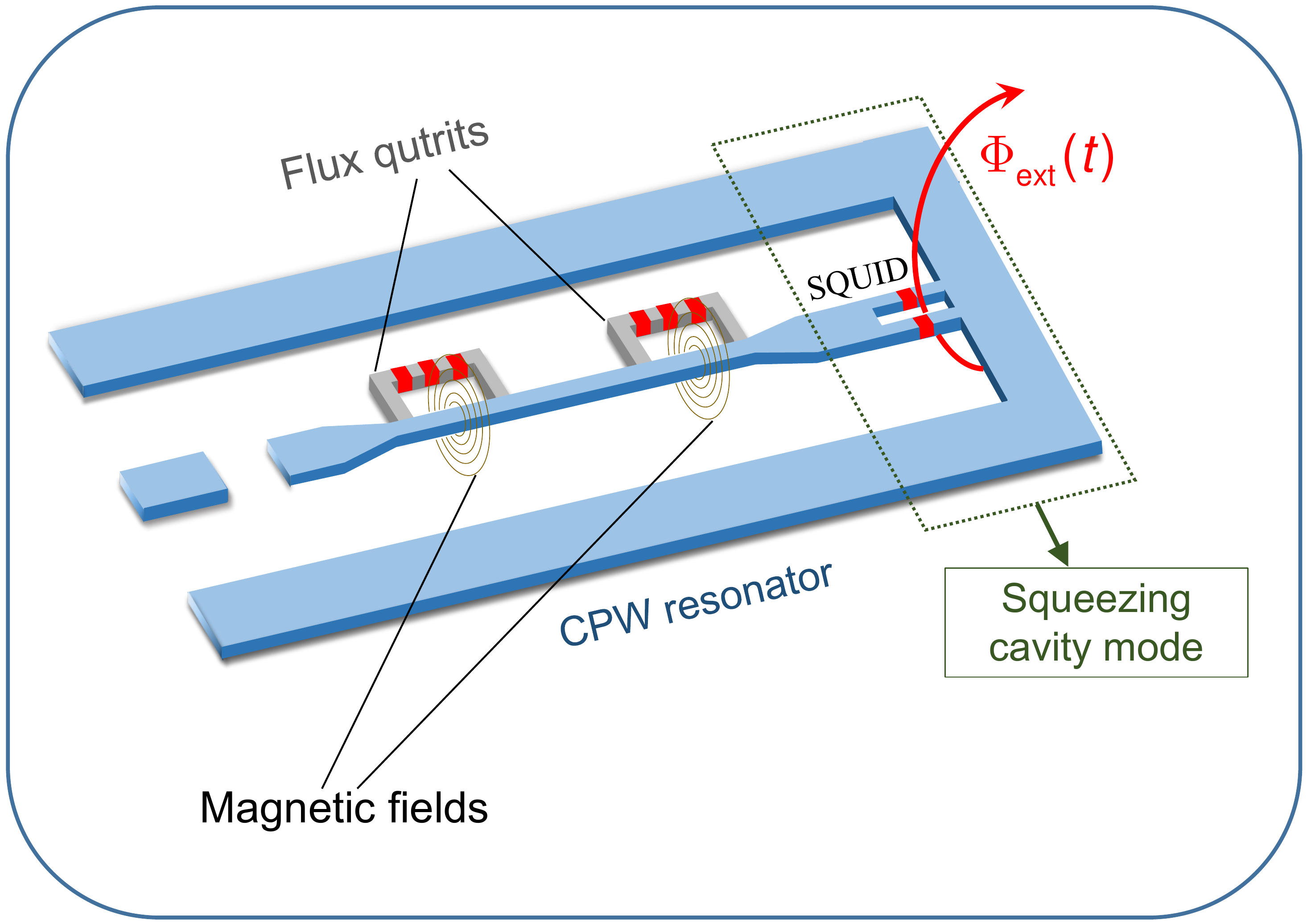}}
	\caption{
		Schematic diagram representing two flux qutrits coupled a coplanar waveguide (CPW) resonator.
		The superconducting quantum interference device (SQUID) controlled by the local magnetic flux $\Phi_{\rm{ext}}(t)$
		threading the loop, creates a squeezed vacuum field in the resonator.
	}
	\label{figend}
\end{figure}

Another alternative system to realize our approach could be superconducting quantum circuits. Figure~\ref{figend}
shows two flux qutrits coupled a coplanar waveguide (CPW) resonator via the induced magnetic field \cite{Np6772,Prl105023601}.
The necessary squeezing in the resonator is created by inserting a superconducting
quantum interference device (SQUID), which is tuned by a magnetic flux $\Phi_{\rm{ext}}(t)$ \cite{Prl103147003,Pra90053833,Rmp841}.
The flux-qubit circuits placed at or near an antinode of the
standing wave of the current on the superconducting wire can strongly
couple to the superconducting resonator via the mutual inductance \cite{Rmp85623}.
The states $|f\rangle$ and $|e\rangle$
correspond to the first and second excited eigenstates of the flux qubit, respectively.
In this system, the transition between $|f\rangle$ and $|g\rangle$ should be much smaller than that between $|e\rangle$ and $|g\rangle$ ($|f\rangle$)
so as to guarantee the final stability of the system.
This is possible to realize by adjusting the magnetic flux \cite{Prb75104516,Prl95087001,Nat4747353,Pr7181} in the superconducting circuit system.

\section{conclusion}
We investigate the possibility of \textit{simultaneously} improving both the evolution speed \textit{and}
the fidelity for a dissipation-based generation of entanglement
by pulse modulation and parametric amplification.
Regarding two typical dissipation sources in this system: atomic spontaneous emission and cavity decay, we
employ atomic spontaneous emission but avoid the effect of cavity decay.
The pulse modulation is used to induce two control functions, $\Xi_{1}(t)$ and $\Xi_{2}(t)$,
where $\Xi_{1}(t)$ is designed to accelerate the population transfer to the target state $|S\rangle$, and
the $\Xi_{2}(t)$ is designed to accelerate the population transfer out of the UDG state $|\psi_{gg}\rangle$.
The parametric amplification is used to increase the cooperativity, and thus to improve the fidelity of the system.
It also allows us to use a relatively large pulse intensity to shorten the stabilization time.

From both analytical and numerical confirmations, we show that the stabilization time in the ADB
is \textit{shortened exponentially} with a controllable squeezing parameter $r_{p}$.
Specifically, when $r_{p}=2$, the stabilization time in ADB approach is 10 times shorter than that in the TDB scheme,
and the fidelity of the ADB approach could reach $98\%$.
We also have analyzed the sensitivity of the speed-up scheme with respect to systematic and stochastic (amplitude-noise) errors.
We find that the ADB approach is robust against parameter errors.
{Therefore, this alternative method can open venues
for the fast and robust realization of high-fidelity entanglement in the presence of dissipation,
and can find wide applications in quantum information technologies.}

\section*{ACKNOWLEDGMENT}
F.N. is supported in part by the:
MURI Center for Dynamic Magneto-Optics via the
Air Force Office of Scientific Research (AFOSR) (FA9550-14-1-0040),
Army Research Office (ARO) (Grant No. W911NF-18-1-0358),
Asian Office of Aerospace Research and Development (AOARD) (Grant No. FA2386-18-1-4045),
Japan Science and Technology Agency (JST) (via the Q-LEAP program, the ImPACT program, and the CREST Grant No. JPMJCR1676),
Japan Society for the Promotion of Science (JSPS) (JSPS-RFBR Grant No. 17-52-50023, and JSPS-FWO Grant No. VS.059.18N),
the RIKEN-AIST Challenge Research Fund, and the
John Templeton Foundation.

\appendix
\section{Derivation of the effective Hamiltonian and the Lindblad operators}

{Squeezing the cavity mode can induce additional noise in the cavity.
A possible strategy to suppress such noise is to introduce a squeezed vacuum field, with a squeezing parameter $r_{e}$ and a reference phase $\theta_{e}$,
to drive the cavity \cite{Prl120093601}. From the point of view of the
cavity, the squeezed input field is well approximated as
having infinite bandwidth \cite{Nat49962}.}
In this case, the dynamical evolution of the system in Fig.~\ref{fig0}
is modeled by a master equation
\begin{align}\label{eq1-a1}
  \dot{\rho}=&i[\rho,H_{0}]+\mathcal{L}_{{a}}\rho+\mathcal{L}_{{c}}\rho, \cr
  \mathcal{L}_{{a}}\rho=&\sum_{k=1}^{4}{\mathcal{L}(L_{k})\rho}, \cr
  \mathcal{L}_{{c}}\rho=&(N+1)\mathcal{L}{(L_{c})}\rho+N\mathcal{L}(L_{c}^{\dag})\rho\cr
                  &-M\mathcal{L}'(L_{c})\rho-M^{*}\mathcal{L}'(L_{c}^{\dag})\rho.
\end{align}
Here, for brevity, we omit the explicit time dependence of $\rho$ and $H_{0}$.
The subscripts ${a}$ and ${c}$ denote the atom and the cavity, respectively.
{The parameter $N$ is the mean photon number of the broadband squeezed field, and $|M|$ determines the
degree of two-photon correlation. The expressions for $N$ and $M$ are \cite{Scully1997,Drummond2004}
\begin{align}
N=&\sinh^{2}(r_{e}), \cr
M=&\cosh(r_{e})\sinh(r_{e})\exp{(-i\theta_{e})},
\end{align}
respectively.} Then, the expressions for $\mathcal{L}(o)\rho$ and $\mathcal{L}'(o)\rho$ are
\begin{align}\label{eq1-a2}
  \mathcal{L}(o)\rho&=o\rho o^{\dag}-\frac{1}{2}(o^{\dag}o\rho+\rho o^{\dag}o), \cr
  \mathcal{L}'(o)\rho&=o\rho o-\frac{1}{2}(oo\rho+\rho oo),
\end{align}
where $o$ denotes the Lindblad operator.
In the system considered here, there are four Lindblad operators $L_{k}$ describing the spontaneous emissions:
\begin{align}\label{eq1-a3}
  L_{1}&=\sqrt{\gamma/2}|f\rangle_{1}\langle e|,\ \ \ L_{2}=\sqrt{\gamma/2}|f\rangle_{2}\langle e|,  \cr
  L_{3}&=\sqrt{\gamma/2}|g\rangle_{1}\langle e|, \ \ \ L_{4}=\sqrt{\gamma/2}|g\rangle_{2}\langle e|,
\end{align}
and one Lindblad operator $L_{c}$ describing the cavity decay:
\begin{align}\label{eq1-a4}
  L_{c}=\sqrt{\kappa}~a.
\end{align}

By introducing the Bogoliubov squeezing transformation $a_{\text{sc}}=\cosh(r_{p})a+e^{-i\theta_{p}}\sinh(r_{p})a^{\dag}$,
we can diagonalize the nonlinear Hamiltonian $H_{\text{NL}}$ as $H_{\text{NL}}=\omega_{\text{sc}}a_{\text{sc}}^{\dag}a_{\text{sc}}$,
where
\begin{align}\label{eq1-a6}
  r_{p}=\frac{1}{4}\ln{\frac{\Omega_{p}+\Delta_{c}}{\Omega_{p}-\Delta_{c}}},
\end{align}
is the squeezing parameter, and $\omega_{\text{sc}}=\sqrt{\Delta_{c}^2-\Omega_{p}^{2}}$ is the squeezed-cavity frequency.
Accordingly, the atom-cavity coupling Hamiltonian $H_{\text{AC}}$ becomes
\begin{align}\label{eq1-a7}
  H_{\text{AC}}=\sum_{j=1,2}(g_{\text{sc}}a_{\text{sc}}-g'_{\text{sc}}a_{\text{sc}}^{\dag})|e\rangle_{j}\langle g|+\text{H.c.},
\end{align}
where $g_{\text{sc}}=g\cosh(r_{p})$ and $g'_{s}=g\exp{(-i\theta_{p})}\sinh{(r_{p})}$.
When $|g'_{\text{sc}}|\ll(\omega_{\text{sc}}+\Delta_{e})$ and $\Delta_{e}=\omega_{\text{sc}}$, the counter-rotating
terms in Eq.~(\ref{eq1-a7}) can be neglected, such that $H_{\text{AC}}$ can be transformed to
\begin{align}\label{eq1-a8}
  H'_{\text{AC}}=g_{\text{sc}}\sum_{j=1,2}{a_{\text{sc}}|e\rangle_{j}\langle g|+\text{H.c.}}.
\end{align}
Meanwhile, the total Hamiltonian in this rotating frame is

\begin{align}\label{eq1-a9}
  H'_{0}=&\sum_{j=1,2}\Omega_{j}(t)|e\rangle_{j}\langle f|+g_{\rm{sc}}|e\rangle_{j}\langle g|a_{\rm{sc}}\cr
         &+\Omega_{j}^{\text{MW}}e^{-i\delta t}|f\rangle_{j}\langle g|+\rm{H.c.}.
\end{align}

\begin{widetext}
The Lindblad term $\mathcal{L}_{\rm{c}}\rho$ in Eq.~(\ref{eq1-a1}) becomes
\begin{align}\label{eq1-a10}
  \mathcal{L}_{\rm{c}}\rho=&(N_{\rm{sc}}+1)\mathcal{L}(L_{\rm{sc}})\rho+N_{\rm{sc}}\mathcal{L}(L_{\rm{sc}}^{\dag})\rho
                  -M_{\rm{sc}}\mathcal{L}'(L_{\rm{sc}})\rho-M_{\rm{sc}^{*}}\mathcal{L}'(L_{\rm{sc}}^{\dag})\rho,
\end{align}
where $L_{\rm{sc}}=\sqrt{\kappa}~a_{\rm{sc}}$ denotes the squeezed-cavity mode decay, $M_{\rm{sc}}$ and $N_{\rm{sc}}$ are
\begin{align}\label{}
  M_{\rm{sc}}=&e^{i\theta_{p}}[\sinh(r_{p})\cosh(r_{e})+e^{-i(\theta_{p}+\theta_{e})}\cosh(r_{p})\sinh(r_{e})]
  \times[\cosh(r_{p})\cosh(r_{e})+e^{i(\theta_{p}+\theta_{e})}\sinh(r_{p})\sinh(r_{e})],\cr
  N_{\rm{sc}}=&\cosh^{2}(r_{p})\sinh^{2}(r_{e})+\sinh^{2}(r_{p})\cosh^{2}(r_{e})
              +\frac{1}{2}\sinh(2r_{p})\sinh(2r_{e})\cos(\theta_{p}+\theta_{e}),
\end{align}
respectively. Then, by choosing $r_{e}=r_{p}$ and $\theta_{e}+\theta_{p}=\pm (2n+1)\pi$ ($n=0,1,2,\ldots$), we obtain
\begin{align}\label{eq1-a11}
  N_{\rm{sc}}=M_{\rm{sc}}=0.
\end{align}
In this case, the master equation in Lindblad form as shown in Eq.~(\ref{eq1-03}) is obtained.

{Assuming that the squeezed cavity is initially in the squeezed vacuum state $|0\rangle_{\rm{sc}}$ and the atoms are initially in their ground states,}
in the limit of $\Omega_{j}(t),\Omega_{j}^{\text{MW}}(t)\ll g_{\rm{sc}}$
\cite{Jmp18756,Pra412295,Prl744763,Pst2149,Prl89080401,Jpa41493001},
the evolution of the system is confined to the effective subspace spanned by
\begin{align}\label{eq1-a12}
  |\psi_{gg}\rangle&=|gg\rangle|0\rangle_{\rm{sc}},\ \ |\psi_{ff}\rangle=|ff\rangle|0\rangle_{\rm{sc}}, \cr
  |T\rangle&=(|fg\rangle+|gf\rangle)|0\rangle_{\rm{sc}}/\sqrt{2}, \cr
  |S\rangle&=(|fg\rangle-|gf\rangle)|0\rangle_{\rm{sc}}/\sqrt{2}, \cr
  |D\rangle&=(|ge\rangle-|eg\rangle)|0\rangle_{\rm{sc}}/\sqrt{2},
\end{align}
where $\{|T\rangle,|S\rangle,|\psi_{gg}\rangle,|\psi_{ff}\rangle\}$ are the basic vectors of the ground-state subspace,
and $|D\rangle$ is the dark state of the excited-state subspace.
{Here, $|0\rangle_{\rm{sc}}$ is a pure single-mode squeezed vacuum state consisting entirely of even-photon Fock state superpositions,
\begin{align}\label{eq1-a12a}
  |0\rangle_{\rm{sc}}=\frac{1}{\sqrt{\cosh{r}}}\sum_{n=0}^{\infty}{(-\tanh r)^{n}\frac{\sqrt{(2n)!}}{2^{n}n!}|2n\rangle}.
\end{align}
The even-photon Fock state $|2n\rangle$ obeys $a|2n\rangle=\sqrt{2n}|2n-1\rangle$ and $a^{\dag}|2n\rangle=\sqrt{2n+1}|2n+1\rangle$.
Single-mode squeezed states are typically generated by degenerate parametric oscillation
in an optical parametric oscillator \cite{Osa41465}, or using four-wave mixing \cite{Prl552409}.}

\begin{figure}
	\centering
	\scalebox{0.28}{\includegraphics{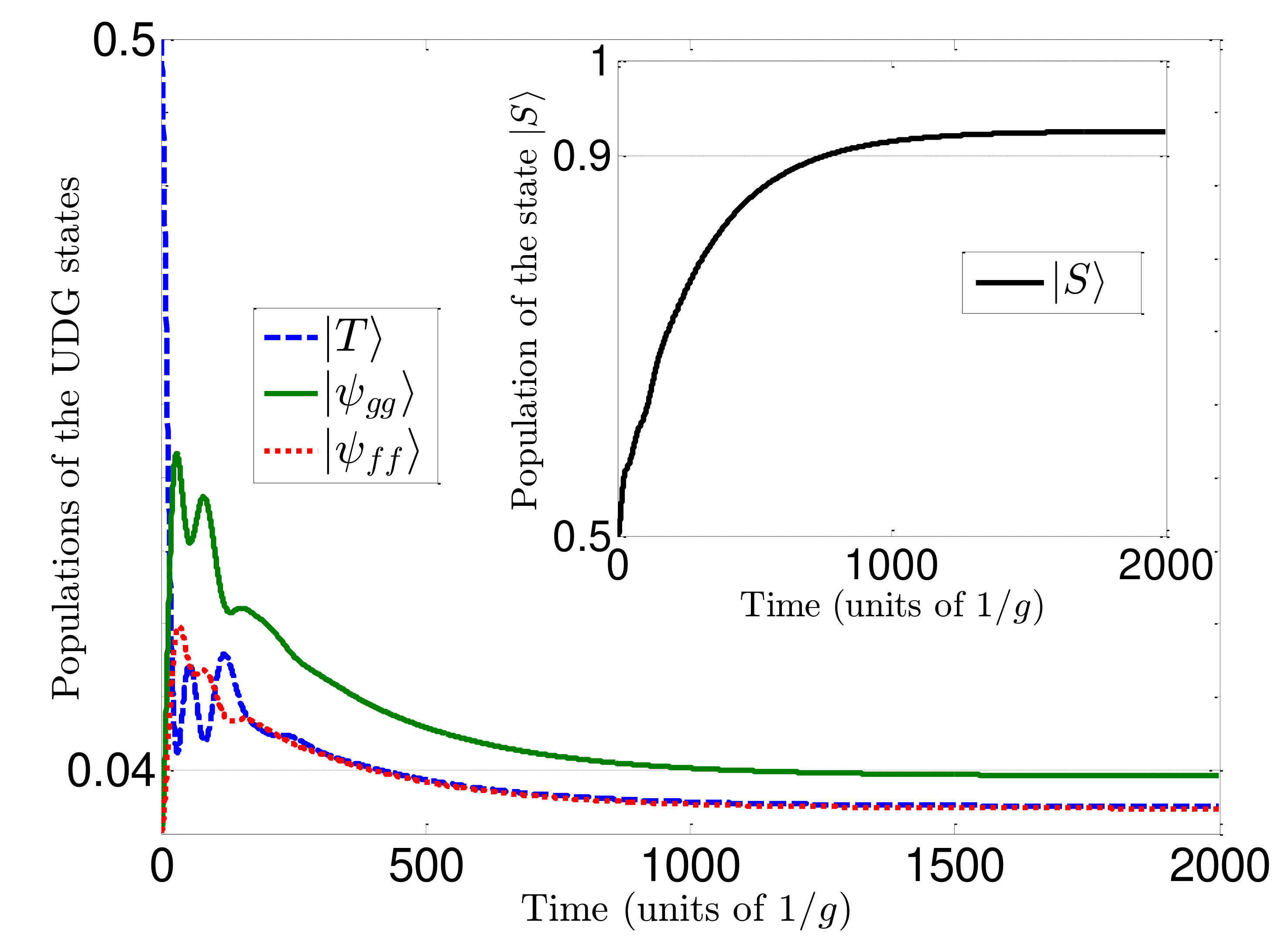}}
	\caption{
		The populations of the UDG states $|T\rangle$, $|\psi_{gg}\rangle$ and $|\psi_{ff}\rangle$ in the TDB method.
		The inset shows the population of the target state $|S\rangle$ versus the evolution time.
		When $t\geq1500/g$, the system becomes stable, i.e., the populations of the ground states gradually become constants.
		The parameters used here are $\Omega_{0}=0.1g$, $\Omega^{\rm{MW}}_{0}=0.2\Omega_{0}$, $\delta=0.4\Omega_{0}$, $\kappa=0.3\gamma$, and $C=g^{2}/(\kappa\gamma)=30$.
	}
	\label{figa1}
\end{figure}

By modulating the Rabi frequencies as
\begin{align}\label{eq1-a13}
  \Omega_{1}(t)=&\Omega_{0}/\sqrt{2}+\Xi_{1}(t),\cr
  \Omega_{2}(t)=&e^{i\pi}\Omega_{0}/\sqrt{2}=\text{constant},\cr
  \Omega_{1}^{\text{MW}}(t)=&\Omega_{0}^{\rm{MW}}/\sqrt{2}+\Xi_{2}(t), \cr
  \Omega_{2}^{\text{MW}}(t)=&\Omega_{0}^{\rm{MW}}\sqrt{2}=\text{constant},
\end{align}
and the effective Hamiltonian for the system becomes
\begin{align}\label{eq1-a14}
  H_{\text{eff}}=&H_{\text{s}}+H_{\text{mod}}, \cr
  H_{\text{s}}=&\frac{1}{\sqrt{2}}[\Omega_{0}| D\rangle+2\Omega_{0}^{\rm{MW}}(e^{i\delta t}|\psi_{gg}\rangle+ e^{-i\delta t}|\psi_{ff}\rangle)]\langle T|+\text{H.c.}, \cr
  H_{\text{mod}}=&\frac{1}{\sqrt{2}}[\Xi_{1}(t)|D\rangle+\Xi_{2}(t)(e^{i\delta t}|\psi_{gg}\rangle+e^{-i\delta t}|\psi_{ff}\rangle)]\langle T|\cr
                 &+\frac{1}{\sqrt{2}}[\Xi_{1}(t)|D\rangle+\Xi_{2}(t)(e^{i\delta t}|\psi_{gg}\rangle- e^{-i\delta t}|\psi_{ff}\rangle)]\langle S|+\text{H.c.}.
\end{align}
\end{widetext}
Here, $H_{\rm{mod}}$ represents the interaction induced by pulse modulation.
Accordingly, we obtain the effective Lindblad operators describing the dissipation processes in the effective evolution subspace as
\begin{align}\label{eq1-a15}
  \tilde{L}_{G}=&\sqrt{\frac{\gamma}{2}}|\psi_{gg}\rangle\langle D|, \ \ \
  \tilde{L}_{T}=\sqrt{\frac{\gamma}{4}}|T\rangle\langle D|, \ \  \ \cr
  \tilde{L}_{S}=&\sqrt{\frac{\gamma}{4}}|S\rangle\langle D|.
\end{align}

According to the effective Hamiltonian in Eq.~(\ref{eq1-a14}) and the effective Lindblad operators in Eq.~(\ref{eq1-a15}),
we find that, $H_{\rm{s}}(\tilde{L}_{k'})|S\rangle=0$ and $\tilde{L}_{k'}^{\dag}|S\rangle=0$ ($k'=G,T,S$).
This means without the pulse modulation ($H_{\rm{mod}}=0$), $|S\rangle$ is a steady state of the system.
Then, the time evolution of the system can be understood as follows: the microwave fields $\Omega_{j}^{\rm{MW}}(t)$
drive the transitions $|\psi_{gg}\rangle\leftrightarrow|T\rangle\leftrightarrow|\psi_{ff}\rangle$,
and the laser fields $\Omega_{j}(t)$ excite $|T\rangle$ to $|D\rangle$, which then decays to $|S\rangle$ via atomic spontaneous emission ($\tilde{L}_{S}$).
In this case, the populations initially in the ground-state subspace are driven to
and trapped in $|S\rangle$, resulting in a maximally entangled state $(|fg\rangle-|gf\rangle)/\sqrt{2}$ (see the inset of Fig.~\ref{figa1}).
Noting that the effective decay rate from $|D\rangle$ to $|\psi_{gg}\rangle$ is two times larger
than those from $|D\rangle$ to $|S\rangle$ and $|T\rangle$, the excited state $|D\rangle$ preferentially decays to the ground state $|\psi_{gg}\rangle$ rather than the other ground states.
As shown in Fig.~\ref{figa1}, the population of the state $|\psi_{gg}\rangle$ increases rapidly to a relatively high level, and then gradually decreases in an oscillating manner.
Meanwhile, the populations of the undesired ground states $|T\rangle$ ($P_{T}$) and $|\psi_{ff}\rangle$ ($P_{ff}$)
decrease quickly to a negligible level.

\bibliography{references}

\end{document}